\def\r#1{\mbox{\bf r}_{#1}}

\documentstyle[multicol,prb,aps,epsf]{revtex}
\begin{document}

\draft
\title{Bound pair states beyond the condensate for Fermi
systems below $T_c\;$:\\the pseudogap as a necessary condition}
\author{A. Yu. Cherny$^{1}$ and A. A. Shanenko$^{2}$}
\address{{}$^{1}$Frank Laboratory of Neutron Physics,
Joint Institute for Nuclear Research,
141980, Dubna, Moscow region, Russia}
\address{{}$^{2}$Bogoliubov Laboratory of Theoretical Physics,
Joint Institute for Nuclear Research,
141980, Dubna, Moscow region, Russia}
\date{November 29, 1998}

\maketitle

\begin{abstract}
As is known, the $1/q^2$ theorem of Bogoliubov asserts that the
mean density of the fermion pair states with the total momentum
$\hbar{\bf q}$ obeys the inequality $n_q \geq C/q^2\ (q \rightarrow
0)$ in the case of the Fermi system taken at nonzero temperature and
in the superconducting state provided the interaction term of its
Hamiltonian is locally gauge invariant.  With the principle of
correlation weakening it is proved in this paper that the reason for
the mentioned singular behaviour of $n_q$ is the presence of the
bound states of particle pairs with nonzero total momenta. Thus,
below the temperature of the superconducting phase transition there
always exist the bound states of the fermion couples beyond the pair
condensate. If the pseudogap observed in the normal phase of the
high--$T_c$ superconductors is stipulated by the presence of the
electron bound pairs, then the derived result suggests,
in a model--independent manner, that the pseudogap survives
below $T_c$.

\end{abstract}
\vspace{0.5cm}

\begin{multicols}{2}

\section{Introduction}

\label{1}

At present the pseudogap is well established to be in the spectrum of the
elementary excitations of undoped and optimally doped high--$T_c$
superconductors~(for example see the review~\cite{Randeria}). The presence of
the pseudogap implies that the electron subsystem in the normal phase is not
the Fermi liquid and, so, theoretical explanation of the pseudogap is
recognized as the key point of understanding the phenomenon of the
high--$T_c$ superconductivity~\cite{Anderson,Mott}.  There are a great number
of various theoretical approaches of investigating this problem. Two of them
considered below are especially interesting in the context of this paper.

The pseudogap can be associated with the presence of the local
pairing correlations without phase coherence. The idea of this
approach assuming the singlet pairing of fermions without the
phase coherence, as applied to the high--$T_c$ superconductivity,
has been proposed in Ref.~\cite{Baskaron}. The more radical model of
Alexandrov and Mott~\cite{Alexandrov} operates with, say, preformed
bosons~(bipolarons) existing in the system above $T_c$, the pseudogap
being treated as coming from the binding energy of a bipolaron (of
the order of a few hundred K). This model dates back to the
Schafroth's ideas according to which the superconductivity is a result
of the Bose--Einstein condensation of the bound pairs of electrons
localized in the space and appearing in the system before the
condensation~\cite{Blatt}.

The concept of a bound state of two particles in medium can
consistently be formulated with the reduced density matrix of the
second order~(2--matrix)~\cite{Bogoliubov1}. Indeed, the system of
two particles is a subsystem of that of $N$ particles. So, its state
is not pure even in the situation when all the system of interest
has a wave function. In general a subsystem is specified by the
density matrix~(see, e.g. Ref.~\cite{Landau}). In particular, the
reduced density matrix of the second order is of use when a
noncoherent superposition of the pure states of two particles is
relevant rather than any wave function. If among these states there
exist bound ones, then a part of particles of the system involved
form bound pair states~\cite{Note1}.  In the superconducting phase a
macroscopical number of particle pairs $N_0$ occupy the same bound
state, i.e. there is the condensate of pairs at which the ratio
$N_0/V=n_0$ is constant in the thermodynamic limit $V \to \infty$.
In the space--uniform case the condensate is formed by the pairs
with the zero total momentum $\hbar{\bf q}=0$, the binding energy
$\varepsilon_{b}$ of these pairs being just the value of the
superconducting gap~\cite{Note9}. The bound particle pairs beyond
the condensate are characterized by the continuous distribution over
the total momentum of a couple~\cite{Note2}. The couple like these
must also have the finite binding energy $\varepsilon_{b}({\bf q})$
that, due to the continuity argument, should tend to
$\varepsilon_{b}$ when $q \to 0$. If these bound particle pairs are
'hard' clusters, like in the theory of Alexandrov and Mott, then one
may consider that the quantity $\varepsilon_{b}({\bf q})$ is
practically independent of ${\bf q}$. The binding energy
$\varepsilon_{b}({\bf q})$ is just the pseudogap, which manifests
itself in the normal phase when the bound couples survive at
$T>T_c$.

In the BCS--theory there are no bound pair states beyond the
condensate absolutely~\cite{Bogoliubov1,Leggett} (see below) which is a
consequence of the violation of the local gauge invariance~(see, for
example, Ref.~\cite{Schrieffer}).

In this paper we shall prove in a model--independent manner
that the existence of the condensate of the bound pair
states (BCS--pairs) implies the presence of the bound couples beyond
the condensate (Schafroth's pairs). Emphasize, that we do not
specify the size of the pairs. If it is much more than the mean 
distance between particles (the condensate pairs in the BCS--model), 
then, following Bogoliubov~\cite{Bogoliubov1}, one may call these 
pairs 'quasi--molecules'. If the radius of the bound particle 
couples is of the order of the mean distance between particles or, 
even, less (the Schafroth--Alexandrov--Mott approach), then one may 
speak about an ordinary molecules. The proof is based on the 
well--known $1/q^2$ theorem of Bogoliubov for the Fermi 
system~\cite{Bogoliubov1} which is valid in the space--uniform case 
and under the condition of the local gauge invariance of the 
interaction term of the system Hamiltonian.

The present article is organized as follows. In section~\ref{2}
the concept of in--medium wave functions of fermion pairs is
considered. The properties of the pair condensate are discussed in 
the third section. At last, the proof concerning the noncondensed 
bound pairs of fermions is given in section~\ref{4} of the paper.

\section{The concept of pair wave functions for fermions}

\label{2}

Thus, let us consider a homogeneous Fermi system of $N$ particles with
the spin $s=1/2$ at nonzero temperatures. Suppose that the total momentum and
spin of the system are conserved quantities. Let the forces
exerted by fermions on each other be described with the two--particle
interaction potential depending on the relative distance between
them and, may be, on the spin variables like in the case of various
effective Hamiltonians. A state of the whole system is specified by
the density matrix corresponding to the canonical Gibbs ensemble:
\begin{equation}
\widehat{\rho}=\exp\biggl(-\frac{\widehat{H}}{k_B T}\biggr)\biggl/
\mbox{Tr}\exp\biggl(-\frac{\widehat{H}}{k_B T}\biggr),
\label{rhoN}
\end{equation}
where $\widehat{H}$ is the system Hamiltonian~\cite{Note3}. In this
case the 2--matrix is represented in the form (see, for example,
\cite{Bogoliubov2})
\begin{eqnarray}
\rho_2(x_1^{\prime},& x_2^{\prime}&; x_1,x_2) \nonumber\\
&&=\frac{1}{N(N-1)}\langle \psi^{\dagger}( x_1) \psi^{\dagger}(x_2)
    \psi (x_2^{\prime})\psi ( x_1^{\prime})\rangle,
\label{rho2}
\end{eqnarray}
where $\langle \cdots \rangle=\mbox{Tr}\,(\cdots \widehat{\rho})$ stands for
the average over the state (\ref{rhoN}); $x=({\bf r},\sigma)$
represents the space coordinates ${\bf r}$ and spin $z$--projection
$\sigma=\pm 1/2$; $\psi^{\dagger}({\bf x}),\,\psi({\bf x})$ are the
field Fermi operators. The 2--matrix obeys the normalization
condition
\begin{equation}
\int dx_1 dx_2\,
\rho_2(x_1, x_2; x_1,x_2)=1,
\label{rho2norm}
\end{equation}
here $\int dx\, \cdots = \sum_\sigma \int d^3r\, \cdots$ and
integration is fulfilled over the volume $V$. Therefore, the
2--matrix (\ref{rho2}) has the asymptotic behaviour $1/V^2$ when
$V\to\infty,\,n=N/V=\mbox{const}$. So, it is more convenient to
deal with the pair correlation function $F_2$ differing by a norm from $\rho_2$:
\begin{equation}
F_2(x_1, x_2; x_1^{\prime}, x_2^{\prime})=
\langle \psi^{\dagger}(x_1) \psi^{\dagger}(x_2)
    \psi (x_2^{\prime})\psi (x_1^{\prime})\rangle .
\label{F2def}
\end{equation}

The boundary conditions for $F_2$~\cite{Note3a} follow from the
principle of the correlation weakening at macroscopical
separations~\cite{Bogoliubov1}:
\begin{eqnarray}
\langle \psi^{\dagger}(x_1) \psi^{\dagger}(x_2)
&\psi&(x_2^{\prime})\psi (x_1^{\prime})\rangle \to\nonumber\\
&&\langle \psi^{\dagger}(x_1) \psi^{\dagger}(x_2)\rangle\;
  \langle\psi (x_2^{\prime})\psi (x_1^{\prime})\rangle
\label{corr1}
\end{eqnarray}
when
\begin{equation}
{\bf r}_1 - {\bf r}_2 = \mbox{const},\
{\bf r}_1^{\prime} - {\bf r}_2^{\prime} = \mbox{const},\
|{\bf r}_1^{\prime} - {\bf r}_1| \to \infty ;
\label{limit1}
\end{equation}
\begin{eqnarray}
\langle \psi^{\dagger}(x_1) \psi^{\dagger}(x_2)
    &\psi&(x_2^{\prime})\psi (x_1^{\prime})\rangle \to \nonumber\\
&&\langle \psi^{\dagger}(x_1) \psi(x_1^{\prime})\rangle\;
  \langle\psi^{\dagger} (x_2)\psi (x_2^{\prime})\rangle
\label{corr2}
\end{eqnarray}
when
\begin{equation}
{\bf r}_1 - {\bf r}_1^{\prime} = \mbox{const},\
       {\bf r}_2 - {\bf r}_2^{\prime} = \mbox{const},\
|{\bf r}_1 - {\bf r}_2| \to \infty.
\label{limit2}
\end{equation}

As the kernel (\ref{F2def}) is a non--negative Hermitian operator
acting on the two--particle wave functions $\psi(x_1,x_2)$, we
can expand it in the orthonormal set of its eigenfunctions (EF):
\begin{equation}
F_{2}(x_1,x_2;x_1^{\prime},x_2^{\prime})=
\sum_{\nu}N_{\nu}\psi_{\nu}^{*}(x_1,x_2)\psi_{\nu}(x_1^{\prime},
x_2^{\prime}),
\label{F2psi}
\end{equation}
where
\begin{equation}
\int dx_1dx_2\,{|\psi_{\nu}
(x_1,x_2)|}^{2}=1.
\label{psinorm}
\end{equation}
Eigenfunctions $\psi_{\nu}(x_1,x_2)$, which at the same time are EF of
2--matrix (\ref{rho2}), are called the pair wave functions, or PWF.

With (\ref{F2def}), (\ref{F2psi}) and (\ref{psinorm}) one can be
convinced that
\begin{eqnarray}
\int dx_1dx_2\,F_{2}(x_1,x_2;x_1,x_2)\;&&=
\langle {\widehat N}^{2}-{\widehat N}\rangle \nonumber\\
&&= N(N-1)=\sum_{\nu}N_{\nu}.
\nonumber
\end{eqnarray}

Therefore, the non--negative quantity $N_{\nu}$ can be interpreted as
the mean number of the pairs in the state $\nu$, any pair being
doubly taken. The ratio $w_{\nu}=N_{\nu}/\{N(N-1)\}$ is the
probability of observing a particle pair in the pure state with the
wave function $\psi_{\nu}(x_1, x_2)$. Here, as one might expect,
$\sum_{\nu} w_{\nu}=1$.

It follows from the definition (\ref{F2def}) that
\begin{eqnarray}
F_{2}(x_1,x_2;x_1^{\prime},x_2^{\prime})\;&&=
-F_{2}(x_1,x_2;x_2^{\prime},x_1^{\prime})=\nonumber\\
&&=-F_{2}(x_2,x_1;x_1^{\prime},x_2^{\prime}).
\nonumber
\end{eqnarray}
So, $\psi_{\nu}(x_1,x_2)=-\psi_{\nu}(x_2,x_1)$, i.e. PWF for
fermions, as usual, are antisymmetric with respect to permutations
of particles.

In an equilibrium state the total pair momentum $\hbar {\bf q}$ is
a good quantum number for PWF provided that the total momentum of
the whole system is a conserving quantity (see proof in
Ref.~\cite{Cherny}). The same is correct for the total spin $S$ of a
particle pair if there is no magnetic ordering~\cite{Note4}.
So, the index $\nu$ can be represented as $\nu=(\omega,{\bf q}, S)$,
where $\omega$ stands for other quantum numbers. As to the PWF, they
can be written as
\begin{eqnarray}
\psi_{\nu}(&&x_1,x_2)\nonumber\\
&&=\psi_{\omega,{\bf q},S}({\bf r}_1-{\bf r}_2,
\sigma_1, \sigma_2)\frac{\exp \{i {\bf q}({\bf r}_1+{\bf r}_2)/2\}}
{\sqrt{V}}.
\label{psiomq}
\end{eqnarray}
Then expression (\ref{F2psi}) has the form
\begin{eqnarray}
&F_{2}&(x_1,x_2;x_1^{\prime},x_2^{\prime})=\sum_{\omega,{\bf q},S}
\frac{N_{\omega,{\bf q},S}}{V}
\psi^{*}_{\omega,{\bf q},S}({\bf r}_1-
                  {\bf r}_2,\sigma_1,\sigma_2)\nonumber\\
&&\times\psi_{\omega,{\bf q},S}(\!{\bf r}_1\!-\!
                               {\bf r}_2,\!\sigma_1\!,\!\sigma_2\!)
\exp\left\{i\frac{{\bf q}}{2}(\!{\bf r}_1^{\prime}\!+
\!{\bf r}_2^{\prime}\!-\!{\bf r}_1\!-\!{\bf r}_2\!)\right\}.
\label{F2psiomq}
\end{eqnarray}

For the wave function $\psi_{\omega,{\bf q},S}({\bf r},\sigma_1,
\sigma_2)$ which can be interpreted as the wave function of
a particle pair in the center--of--mass system, from (\ref{psinorm})
and (\ref{psiomq}) we obtain
\begin{equation}
\sum\limits_{\sigma_1,\sigma_2} \int\limits_V\,d^3r
|\psi_{\omega,{\bf q},S}({\bf r},\sigma_1,\sigma_2)|^2=1.
\label{psiomqnorm}
\end{equation}
It can be related to either discrete or continuous spectra. In the
former case
\begin{equation}
\psi_{\omega,{\bf q},S}({\bf r},\sigma_1,\sigma_2) \to 0
\label{limdiscr}
\end{equation}
when $r \to \infty$ and, so, we deal with the sector of bound
states of particle pairs. The latter variant implies
\begin{equation}
\psi_{\omega,{\bf q},S}({\bf r},\sigma_1,\sigma_2) \to
\chi_{S,m_S}(\sigma_1,\sigma_2)
\sqrt{2}\left\{\begin{array}{c}
\cos({\bf p}{\bf r})\\
\sin({\bf p}{\bf r})
\end{array}\right.
\label{limcont}
\end{equation}
for $r \to \infty$. This is a 'dissociated', or scattering, pair
state corresponding to the relative motion with the momentum
$\hbar{\bf p}$. Here $\chi_{S,m_S}(\sigma_1,\sigma_2)$ is the spin
part of the pair wave function (spinor), $m_S$ being the $z$--projection of
the total pair spin $S$. When $S=0$, $m_S=0$ (the singlet state)
one should take $\cos({\bf p}{\bf r})$. For $S=1$, $m_S=-1,0,1$ (the triplet
state) one should use $\sin({\bf p}{\bf r})$. Remark
that in the situation when the fermion interaction does not depend
on spin variables, the spin and space parts of the PWF can be
separated from one another not only when $r \to \infty$ but also
for any ${\bf r}$.

In the case of (\ref{limdiscr}) $\omega=i$, where $i$ stands for
the discrete index enumerating the bound pair states. Let us
denote $\psi_{\omega,{\bf q},S}({\bf r},\sigma_1,\sigma_2)=
\varphi_{{\bf q},S,i}({\bf r},\sigma_1,\sigma_2)$, so that
\begin{equation}
\sum\limits_{\sigma_1,\sigma_2} \int\limits_V d^3r\,
|\varphi_{i,{\bf q},S}({\bf r},\sigma_1,\sigma_2)|^2=1.
\label{phibnorm}
\end{equation}
In the situation of (\ref{limcont}) $\omega=({\bf p},m_S)$.
Here it is convenient to introduce
$\psi_{\omega,{\bf q},S}({\bf r},\sigma_1,\sigma_2)=
\varphi_{{\bf p},{\bf q},S,m_S}({\bf r},\sigma_1,\sigma_2)/\sqrt{V}$.
From (\ref{psiomqnorm}) it follows that
\begin{equation}
\frac{1}{V}\sum\limits_{\sigma_1,\sigma_2} \int\limits_V d^3r\,
|\varphi_{{\bf p},{\bf q},S,m_S}({\bf r},\sigma_1,\sigma_2)|^2=1.
\label{phicnorm}
\end{equation}
Now, with the variables
\begin{equation}
{\bf R}=(\r{1}+\r{2})/2,\quad {\bf r}=\r{1}-\r{2}
\label{Rr}
\end{equation}
and, respectively, ${\bf R}^{\prime}$ and ${\bf r}^{\prime}$, the
expression (\ref{F2psiomq}) is rewritten as
\begin{eqnarray}
&F_{2}&(x_1,x_2;x_1^{\prime},x_2^{\prime})=
\sum_{{\bf q},S,i}\frac{N_{{\bf q},S,i}}{V}
\varphi^{*}_{{\bf q},S,i}({\bf r},\sigma_1,\sigma_2)\nonumber\\
&&\times\varphi_{{\bf q},S,i}
({\bf r}^{\prime},\sigma_1^{\prime},\sigma_2^{\prime})
\exp\left\{i{\bf q}({\bf R}^{\prime}-{\bf R})\right\}
\nonumber\\
&&+\!\sum_{{\bf p},{\bf q},S,m_S}\!\!
                             \frac{N_{{\bf p},{\bf q},S,m_S}}{V^2}
\varphi^{*}_{{\bf p},{\bf q},S,m_S}({\bf r},\sigma_1,\sigma_2)
\nonumber\\
&&\times\varphi_{{\bf p},{\bf q},S,m_S}({\bf r}^{\prime},
 \sigma_1^{\prime},\sigma_2^{\prime})\exp\left\{i
       {\bf q}({\bf R}^{\prime}-{\bf R})\right\}.
\label{F2discr}
\end{eqnarray}
In the thermodynamic limit all the summations over momenta can be
replaced by the corresponding integrals:
\begin{eqnarray}
&F_{2}&(x_1,x_2;x_1^{\prime},x_2^{\prime})=
\sum_{S,i}\int d^3q\,w_{S,i}({\bf q})\;
\varphi^{*}_{{\bf q},S,i}({\bf r},\sigma_1,\sigma_2)\nonumber\\
&&\times\varphi_{{\bf q},S,i}
({\bf r}^{\prime},\sigma_1^{\prime},\sigma_2^{\prime})
\exp\left\{i{\bf q}({\bf R}^{\prime}-{\bf R})\right\}
\nonumber\\
&&+\!\sum_{S,m_S}\!\!\int\!d^3p\,d^3q \,w_{S,m_S}({\bf p},{\bf q})\;
\varphi^{*}_{{\bf p},{\bf q},S,m_S}({\bf r},\sigma_1,\sigma_2)
\nonumber\\
&&\times\varphi_{{\bf p},{\bf q},S,m_S}({\bf r}^{\prime},
 \sigma_1^{\prime},\sigma_2^{\prime})
\exp\left\{i{\bf q}({\bf R}^{\prime}-{\bf R})\right\}.
\label{F2cont}
\end{eqnarray}

Thus, from equations (\ref{F2discr}) and (\ref{F2cont}) we can see
that $V\,w_{S,i}({\bf q})d^3q$ is the number of the bound particle pairs
with the spin $S$, in the state $i$ and with the total couple
momentum $\hbar {\bf q}$ located in the infinitesimal volume $d^3q$.
Respectively, $V^2\,w_{S,m_S}({\bf p},{\bf q})d^3p\,d^3q$ stands for
the number of the 'dissociated' particle pairs in the state $(S,m_S)$
with the relative momentum $\hbar {\bf p}$ and total momentum $\hbar
{\bf q}$ located in the infinitely small volumes $d^3p$ and $d^3q$.

In the center--of--mass system the replacement ${\bf p} \to -
{\bf p},\; \sigma_1 \to \sigma_2,\; \sigma_2 \to \sigma_1$
corresponds to the permutation of particles. So,
the following symmetric relations take place:
\begin{equation}
w_{S,m_S}({\bf p},{\bf q})=w_{S,m_S}(-{\bf p},{\bf q}),
\label{wpqsimm}
\end{equation}
\begin{eqnarray}
\varphi_{{\bf p},{\bf q},S,m_S}({\bf r},
 \sigma_1,\sigma_2)=&& \,-\,
\varphi_{{\bf p},{\bf q},S,m_S}(-{\bf r},
 \sigma_2,\sigma_1)\nonumber\\
=&&\,-\,\varphi_{-{\bf p},{\bf q},S,m_S}({\bf r},
 \sigma_2,\sigma_1).
\label{phipqsim}
\end{eqnarray}

As an example, let us consider the expansion of $F_2$ in terms
of PWF for the BCS--model. Taken with an accuracy to the asymptotically
small quantities, the Hamiltonian in the BCS--approach is represented
as the quadratic form of the Fermi operators~\cite{Bogoliubov3} that can be
diagonalized with the Bogoliubov transformation.
Therefore, one is able to use the theorem of Wick, Bloch and
De Dominicis~\cite{Bloch}:
\begin{eqnarray}
&&F_2(x_1, x_2; x_1^{\prime}, x_2^{\prime})=
\langle \psi^{\dagger}(x_1) \psi^{\dagger}(x_2)
 \psi (x_2^{\prime})\psi (x_1^{\prime})\rangle \nonumber\\[2mm]
&&=\langle \psi^{\dagger}(x_1) \psi^{\dagger}(x_2)\rangle\,
 \langle \psi (x_2^{\prime})\psi (x_1^{\prime})\rangle +
\langle \psi^{\dagger}(x_1) \psi(x_1^{\prime})\rangle\,\nonumber\\ [2mm]
&&\times\langle \psi^{\dagger}(x_2) \psi (x_2^{\prime})\rangle
-\langle \psi^{\dagger}(x_1) \psi(x_2^{\prime})\rangle\,
 \langle \psi^{\dagger} (x_2)\psi (x_1^{\prime})\rangle .
\label{F2wick}
\end{eqnarray}
Further, for the 'normal' averages we have
\begin{eqnarray}
&&\langle \psi^{\dagger}(x_1) \psi(x_1^{\prime})\rangle=
\langle \psi^{\dagger}({\bf r}_1,\sigma_1) \psi({\bf r}_1^{\prime},
\sigma_1^{\prime})\rangle\nonumber\\
&&=\int\frac{d^3k}{(2\pi^3)}n(k)
\exp\left\{i\,{\bf k}\,({\bf r}_1^{\prime}-{\bf r}_1)\right\}\,
\Delta(\sigma_1-\sigma_1^{\prime}),
\label{F1}
\end{eqnarray}
where $n(k)=\langle a^{\dagger}_{{\bf k},\sigma} a_{{\bf k},\sigma}\rangle$
gives the distribution of fermions over momenta; and we introduced the
function
$$
\Delta(\sigma)=\left\{\begin{array}{ll}
0, &\sigma\not=0, \\
1, &\sigma=0.
\end{array}\right.
$$
'Anomalous' averages are given by
\begin{eqnarray}
&& \langle \psi(x_1) \psi(x_1^{\prime})\rangle=
\langle \psi({\bf r}_1,\sigma_1) \psi({\bf r}_1^{\prime},
\sigma_1^{\prime})\rangle=\nonumber\\
&&=\!\!\!\int\!\!\!\frac{d^3k}{(2\pi)^3}\langle a_{{\bf k},\sigma_1}
a_{-{\bf k},-\sigma_1}\rangle
\exp\{\! i{\bf k}({\bf r}_1^{\prime}-{\bf r}_1)\}
\Delta(\sigma_1\!+\!\sigma_1^{\prime}).
\label{F1Anom}
\end{eqnarray}
In the BCS--model, the quantity $\langle a_{{\bf k},
\sigma} a_{-{\bf k},-\sigma}\rangle$ can be represented in the
following form
\begin{equation}
\langle a_{{\bf k},\sigma} a_{-{\bf k},-\sigma}\rangle =
\sqrt{n_0} \;\varphi(k)\; \frac{\mbox{sign}(\sigma)}{\sqrt{2}},
\label{akak}
\end{equation}
with $\varphi(k)$ obeying the normalization condition
\begin{equation}
\int\frac{d^3 k}{(2\pi)^3}\;|\varphi(k)|^2 =1.
\label{phinorm}
\end{equation}
Remark that one can consider $\varphi(k)$ as a real quantity
because it can be made real with the corresponding phase
transformation of the operators $a_{\bf k}$ and $a^{\dagger}_{\bf k}$.
Now, Eqs. (\ref{F1}), (\ref{F1Anom}) and
(\ref{akak}) allow us to rewrite (\ref{F2wick}) in the following
form:
\begin{eqnarray}
&&F_2(x_1, x_2; x_1^{\prime}, x_2^{\prime})=
n_0\varphi(r) \chi_{0,0}(\sigma_1,\sigma_2)\varphi(r^{\prime})
\chi_{0,0}(\sigma_1^{\prime},\sigma_2^{\prime})\nonumber\\[1mm]
&&+\!\!\sum\limits_{S,m_S}\!\!\int \!\!\frac{d^3 pd^3 q}{(2\pi)^6}
n\Bigl(\frac{{\bf q}}{2}+{\bf p}\Bigr) n\Bigl(\frac{{\bf q}}{2}-{\bf p}\Bigr)
\varphi_{{\bf p}, S}({\bf r}) \chi_{S,m_S}(\sigma_1,\sigma_2)
\nonumber\\[1mm]
&&\times \varphi_{{\bf p}, S}({\bf r}^{\prime})
\chi_{S,m_S}(\sigma_1^{\prime},\sigma_2^{\prime})
\exp\left\{i {\bf q} ({\bf R}^{\prime}-{\bf R})\right\}.
\label{F2BCS}
\end{eqnarray}
Here $\varphi(r)$ is the Fourier transform of $\varphi(k)$,
for $\varphi_{{\bf p},S}({\bf r})$ we have
\begin{equation}
\varphi_{{\bf p},S}({\bf r})=\left\{\begin{array}{ll}
\sqrt{2}\cos({\bf p}{\bf r}), &S=0,\\
\sqrt{2}\sin({\bf p}{\bf r}), &S=1.
\end{array}\right.
\label{phiplane}
\end{equation}
Respectively, the spinor $\chi$ stands for
\begin{eqnarray}
\chi&&_{S,m_S}(\sigma_1,\sigma_2)\nonumber\\[1mm]
    &&=\left\{\begin{array}{ll}
\Delta(\sigma_1+\sigma_2) \mbox{sign}(\sigma_1)/\sqrt{2}, &S=0,\,m_S=0;\\[1mm]
\Theta(-\sigma_1) \Theta(-\sigma_2),              &S=1,\,m_S=-1;\\[1mm]
\Delta(\sigma_1+\sigma_2)/\sqrt{2},               &S=1,\,m_S=0;\\[1mm]
\Theta(\sigma_1) \Theta(\sigma_2),                &S=1,\,m_S=1.
\end{array}\right.
\label{chisms}
\end{eqnarray}
Here
$$
\Theta(\sigma)
=\left\{\begin{array}{ll}
1,& \sigma \geq 0,\\
0,& \sigma < 0.
\end{array}\right.
$$
With (\ref{phinorm}), (\ref{phiplane}) and (\ref{chisms}) one can
easily be convinced that the normalization relations (\ref{phibnorm})
and (\ref{phicnorm}) are satisfied. Within the BCS--model
$w_{S,i}({\bf q})=\Delta(S)\Delta(i)\,n_0\,\delta({\bf q})$\
($\delta({\bf q})$) is the $\delta$-function), i.e. all the bound
particle pairs are condensed.

\section{Properties of the condensate of pairs}

\label{3}

Let us demonstrate in the most general case that if the 'anomalous'
average $\langle\psi(x_1)\psi(x_2)\rangle$ is not equal to zero
(off diagonal long--range order) then the distribution function
$w_{S,i}({\bf q})$ acquires the $\delta-$functional singularity
corresponding to some indices $S_0$ and $i_0$ or, in other words,
the ratio $N_{{\bf q}, S_0, i_0}/V$ in the first sum of
(\ref{F2discr}) does not vanish in the thermodynamic limit:
\begin{equation}
w_{S,i}({\bf q})=n_0\delta({\bf q})\Delta(S-S_0)\Delta(i-i_0)
+ \widetilde w_{S,i}({\bf q}),
\label{wiq}
\end{equation}
where $\widetilde w_{S,i}({\bf q})$ is the regular part of
(\ref{wiq}) giving the bound--pair distribution over nonzero
momenta.

To do this, let us take the limit relation (\ref{corr1}) and rewrite
it with the variables (\ref{Rr}) in the form
\begin{eqnarray}
F_2(&x_1&, x_2; x_1^{\prime}, x_2^{\prime}) \to
\langle \psi^{\dagger}(x_1) \psi^{\dagger}(x_2)\rangle\;\nonumber\\
&&\times\langle\psi (x_2^{\prime})\psi (x_1^{\prime})\rangle=
n_0\varphi^*({\bf r},\sigma_1,\sigma_2)
      \varphi({\bf r}^{\prime},\sigma_1^{\prime},\sigma_2^{\prime}),
\label{n0phiphi}
\end{eqnarray}
where the functions $\varphi^*({\bf r},\sigma_1,\sigma_2)$
and $\varphi({\bf r}^{\prime},\sigma_1^{\prime},\sigma_2^{\prime})$
are introduced in such a way that the normalization condition
(\ref{phibnorm}) should be fulfilled. This can always be done
because according to the principle of correlation
weakening~\cite{Bogoliubov1}
$$
\langle\psi (x_1)\psi (x_2)\rangle \to
\langle\psi (x_1)\rangle\;\langle\psi (x_2)\rangle = 0
$$
when $r \to \infty$ (see Ref.~\cite{Note5}). Expression (\ref{n0phiphi})
is exactly the contribution of the first singular term of
(\ref{wiq}) into (\ref{F2cont}). The contribution of the regular
part of (\ref{wiq}) and that of the 'dissociated' pair states into
(\ref{F2cont}) are infinitely small in the situation of
(\ref{limit1}) according to the Riemann's theorem~\cite{Note6}
because
$$
|{\bf R}^{\prime}-{\bf R}| = \Bigl|\frac{{\bf r}_1^{\prime}+{\bf
r}_2^{\prime}}{2}-\frac{{\bf r}_1+{\bf r}_2}{2}\Bigr| \to \infty.
$$

Remark that the pair distribution over the scattering states
(\ref{wpqsimm}) does not contain $\delta-$functional terms.
Indeed, in the opposite case they would lead to the condensate of
the one--particle states like in the situation of the Bose
liquid~\cite{Cherny} which is impossible for the Fermi systems.

Eq. (\ref{n0phiphi}) allows us to treat the 'anomalous'
averages as the wave functions of the condensed pairs of fermions,
(of course, with an accuracy to the normalizing factor). For the
density of the pairs like these Eq.(\ref{phibnorm}) and
(\ref{n0phiphi}) gives
\begin{eqnarray}
n_0&=&\frac{N_{{\bf q}=0,S_0,i_0}}{V}=
   \sum\limits_{\sigma_1,\sigma_2}\int d^3 r\,|\langle \psi({\bf r},
                  \sigma_1)\psi(0, \sigma_2)\rangle|^2=\nonumber\\
&&=\sum\limits_{\sigma}\int \frac{d^3 k}{(2\pi)^3}
       |\langle a_{{\bf k},\sigma} a_{-{\bf k},-\sigma}\rangle|^2,
\label{n0}
\end{eqnarray}
where it has been taken into account that the total momentum of the
system and $z$--component of its spin are conserved quantities.
Keeping in mind these integrals of the motion, one could expect that
in the most general case the wave function of the condensed pairs should
be written as
\begin{eqnarray}
&&\varphi({\bf r},\sigma_1,\sigma_2)=\frac{1}{\sqrt{n_0}}
\langle \psi({\bf r}_1,\sigma_1)\psi({\bf r}_2,
          \sigma_2)\rangle\nonumber\\
&&=\int \frac{d^3 k}{(2\pi)^3}
  \langle a_{{\bf k},\sigma_1} a_{-{\bf k},-\sigma_1}\rangle
\frac{\Delta(\sigma_1+\sigma_2)}{\sqrt{n_0}}
         \exp(i{\bf k} {\bf r})\nonumber\\
&&=\frac{1}{\sqrt{n_0}}\frac{\Delta(\sigma_1+\sigma_2)}{\sqrt{2}}
(\varphi_s({\bf r}) \mbox{sign}(\sigma_1)+ \varphi_t({\bf r})),
\label{phicond}
\end{eqnarray}
where $\varphi_s({\bf r})=\varphi_s(-{\bf r})$ and $\varphi_t({\bf
r})=-\varphi_t(-{\bf r})$. According to Eq. (\ref{chisms}) the first
term in (\ref{phicond}) corresponds to the singlet and the second,
to the triplet components of the wave function of the condensed
fermions. However, (\ref{phicond}) is not quite correct because the
total pair spin should be an integral of the motion, even in the
situation with the spin--dependent interaction between fermions.
Therefore, we are not able to obtain a superposition of the singlet
and triplet states. Instead, in (\ref{phicond}) one should select
either $\varphi_s({\bf r})\not=0$, $\varphi_t({\bf r})=0$ or
$\varphi_s({\bf r})=0$, $\varphi_t({\bf r})\not=0$. So, we have:
\begin{equation}
\varphi({\bf r}, \sigma_1,\sigma_2)=\left\{
\begin{array}{l}
\varphi_s({\bf r}) \chi_{0,0}(\sigma_1,\sigma_2)/\sqrt{n_0},\\[2mm]
\varphi_t({\bf r}) \chi_{1,0}(\sigma_1,\sigma_2)/\sqrt{n_0}.\\
\end{array}
\right.
\label{phiconda}
\end{equation}

The phase coherence takes place for the condensed bound pairs
due to the uncertainty relation $\Delta \varphi \Delta N_0 \simeq
1$ for the phase $\varphi$ and number of the bound fermion pairs
$N_0=N_{{\bf q}=0,S_0,i_0}$ in the state $({\bf q}=0,S_0,i_0)$.
In the thermodynamic limit the macroscopical occupation of this
state results in $\Delta N_0 \propto \sqrt{N_0} \to \infty$ and,
therefore, $\Delta \varphi \to 0$. For the bound pair states
beyond the condensate $N_{{\bf q}, S, i}$ is limited above even
for $V \to \infty$. Thus these states are not correlated with
respect to the phase.

Remark that the total number of the bound particle pairs (condensed
and not)
$$N_b=V\sum\limits_{S,i}\int d^3 q \,w_{S,i}({\bf q})
$$
is proportional to the total number of particles $N$. In particular,
there is the inequality for the number of the condensed bound
pair states~\cite{Yang}
\begin{equation}
N_0 \leq N.
\label{N0N}
\end{equation}
It should be emphasized that the inequality (\ref{N0N}) is not trivial.
One can consider, for example, a dilute gas of $m$--particle molecules.
In this case we have $N_{b}=(m-1)N$, thus, one can obtain $N_{b}>N$ provided
$m\ge 3$.

In the space--uniform case we can readily find
relation (\ref{N0N}) with the inequality of
Cauchy--Schwarz--Bogoliubov~\cite{Bogoliubov1}
$$
|\langle \widehat{A}\widehat{B}\rangle|^{2} \leq
    \langle\widehat{A}\widehat{A}^{\dagger}\rangle
    \langle\widehat{B}^{\dagger}\widehat{B}\rangle.
$$
Indeed, assuming $A=a_{{\bf k}, \sigma}$ and $B=a_{-{\bf k},
-\sigma}$ we arrive at
\begin{eqnarray}
|\langle a_{{\bf k},\sigma}\,a_{-{\bf k},-\sigma}\rangle|^2 \leq
\langle a_{{\bf k},\sigma}a_{{\bf k},\sigma}^{\dagger}\rangle\,
&&\langle a_{-{\bf k},-\sigma}^{\dagger} \,a_{-{\bf k},-\sigma}
\rangle\nonumber\\
&&=\left(1-n(k)\right)\,n(k).
\nonumber
\end{eqnarray}
Then, from (\ref{n0}) we derive
\begin{eqnarray}
&&n_0=\frac{N_0}{V}=\frac{1}{V}\sum\limits_{{\bf k},\sigma}
  |\langle a_{{\bf k},\sigma} a_{-{\bf k},-\sigma}\rangle|^2
\nonumber\\
&&\leq \frac{2}{V}\sum_{{\bf k}}\,\left(n(k)-n^2(k)\right) \leq
\frac{2}{V}\sum_{{\bf k}}n(k)=\frac{N}{V}=n.
\nonumber
\end{eqnarray}
It is interesting to note that $n(k)-n^2(k)= \langle (a^{\dagger}_{{\bf
k},\sigma}a_{{\bf k},\sigma})^2\rangle-\langle a_{{\bf k},\sigma}^{\dagger}
a_{{\bf k},\sigma}\rangle^2=D\bigl(n(k)\bigr)$ is nothing else but the mean
square deviation of the occupation number of the $({\bf k}, \sigma)$
one--particle state. So, the stronger inequality
\begin{equation}
n_0 \leq\frac{2}{V}\sum\limits_{{\bf k}}D\bigl(n(k)\bigr)
\label{N0Dnk}
\end{equation}
demonstrates that the
number of the condensed pairs is tightly connected with the 'wash--out' of
the Fermi surface. In the BCS--model at zero temperature
$$
\frac{n_0}{n} \propto \frac{k_{B}T_c}{E_F} \ll 1
$$
because the bound pairs are formed by the particles located
near the Fermi surface only. In general, $n_0$ is the most 'reliable'
order parameter of the superconducting phase transition.

\section{The Bogoliubov $1/q^{2}$ theorem and bound pair states beyond
the condensate}

\label{4}

Let us now prove with the principle of the correlation weakening
that the distribution of the particle pairs over the 'scattering'
states $w_{S,m_S}({\bf p},{\bf q})$ is expressed in terms of the
occupation numbers of the one--particle states
$n(k)=\langle a^{\dagger}_{{\bf k},\sigma}a_{{\bf k},\sigma} \rangle$.
Indeed, on the one hand, in the limiting situation of (\ref{limit2}) we have
the relation (\ref{corr2}), which can be written as
\begin{eqnarray}
&&F_2(x_1,x_2;x_1^{\prime},x_2^{\prime})
\to \nonumber\\[1mm]
&&\int\frac{d^3 p_1}{(2\pi)^3} n(p_1) \exp(i{\bf p}_1
({\bf r}_1^{\prime}-{\bf r}_1)) \Delta(\sigma_1-\sigma_1^{\prime})
\nonumber\\[1mm]
&&\times\int\frac{d^3 p_2}{(2\pi)^3} n(p_2) \exp(i{\bf p}_2
({\bf r}_2^{\prime}-{\bf r}_2)) \Delta(\sigma_2-\sigma_2^{\prime})
\nonumber\\[1mm]
&&=\int d^3q\,d^3p \frac{n({\bf q}/2+{\bf p})\,
         n({\bf q}/2-{\bf p})}{(2\pi)^6}
\exp(i{\bf p}({\bf r}^{\prime}-{\bf r}))\nonumber\\[1mm]
&&\times \exp(i{\bf q}({\bf R}^{\prime}
-{\bf R})) \Delta(\sigma_1-\sigma_1^{\prime})
                       \Delta(\sigma_2-\sigma_2^{\prime}),
\label{F1F1}
\end{eqnarray}
where, passing to the last equality, we introduced the new
variables ${\bf q}={\bf p}_1+{\bf p}_2$ and ${\bf p}=({\bf p}_1-
{\bf p}_2)/2$ and used notations (\ref{Rr}). On the other hand,
when (\ref{limit2}) is true, we have
$$r=|{\bf r}_2-{\bf r}_1| \to \infty,\
   r^{\prime}=|{\bf r}_2^{\prime}-{\bf r}_1^{\prime}|\to \infty,
$$
$$
|{\bf r}+{\bf r}^{\prime}| \to \infty,\ {\bf R}^{\prime}-{\bf R}
=\mbox{const},\ {\bf r}^{\prime}-{\bf r}=\mbox{const}.
$$
Therefore, it follows from (\ref{limdiscr}), (\ref{limcont}) and
(\ref{F2cont}) that in the limiting case $(\ref{limit2})$ we have
\begin{eqnarray}
&&F_2(x_1,x_2;x_1^{\prime},x_2^{\prime}) \to \nonumber\\
&&\sum\limits_{S,m_S}\!\!\int d^3 q\,d^3 p\; w_{S,m_S}({\bf
p},{\bf q})\varphi_{{\bf p}, S}({\bf r})\varphi_{{\bf p}, S}
({\bf r}^{\prime}) \nonumber\\
&&\times\chi_{S,m_S}(\sigma_1,\sigma_2) \chi_{S,m_S}
(\sigma_1^{\prime},\sigma_2^{\prime})
\exp\left(i{\bf q}({\bf R}^{\prime}-{\bf R})\right),
\label{F2lim}
\end{eqnarray}
where we used notations (\ref{phiplane}) and (\ref{chisms}).
Further, the Riemann's theorem~\cite{Note6} used while
integrating over ${\bf p}$ and relation (\ref{wpqsimm}) allow us
to rewrite (\ref{F2lim}) as
\begin{eqnarray}
&&F_2(x_1,x_2;x_1^{\prime},x_2^{\prime}) \to \nonumber\\
&&\int d^3q\,d^3p \sum\limits_{S,m_S}\; w_{S,m_S}({\bf
p},{\bf q})\chi_{S,m_S}(\sigma_1,\sigma_2) \nonumber\\
&&\times\chi_{S,m_S}(\sigma_1^{\prime},\sigma_2^{\prime})
\exp\left(i{\bf p}({\bf r}^{\prime}-{\bf r})\right)
\exp\left(i{\bf q}({\bf R}^{\prime}-{\bf R})\right).
\label{F2lim2a}
\end{eqnarray}
The right--hand side of Eq. (\ref{F1F1}) is equal to that of
(\ref{F2lim2a}) at all the values of the spin variables and space
ones $\tilde{\bf r}={\bf r}^{\prime}-{\bf r}$ and 
$\tilde{\bf R}={\bf R}^{\prime}-{\bf R}$. 
Taking into account the completeness of the set of
the spin functions (\ref{chisms})
$$
\!\sum\limits_{S,m_S}\!\!\chi_{S,m_S}(\sigma_1,\sigma_2)
 \chi_{S,m_S}(\sigma_1^{\prime},\sigma_2^{\prime})=\Delta(\sigma_1
-\sigma_1^{\prime})\Delta(\sigma_2-\sigma_2^{\prime}),
$$
we derive the following equality:
\begin{equation}
w_{S,m_S}({\bf p},{\bf q})=\frac{n({\bf q}/2+{\bf p})\,
                    n({\bf q}/2-{\bf p})}{(2\pi)^6}.
\label{wpqsms}
\end{equation}
Thus, in the thermodynamic limit one can
write
\begin{equation}
N_{{\bf p},{\bf q},S,m_S}=n({\bf q}/2+{\bf p})\,
                    n({\bf q}/2-{\bf p}).
\label{Npqsms}
\end{equation}
As it is seen, when there is no magnetic ordering (it is obviously
true for the superconducting phase), the function of the pair
distribution over the 'dissociated' states is independent of the
quantum numbers $S,m_S$.

It is now easy to prove that the pair condensate must always
be accompanied by the presence of the noncondensed fermion pairs:
$\widetilde w_{S,i}({\bf q})\not=0$ in (\ref{wiq}) if $n_0\not=0$.
Let the interaction energy of the system be invariant with respect
to the local gauge transformation of the field Fermi
operators~\cite{Note7}
\begin{eqnarray}
\psi({\bf r},\sigma)           &\to& \psi({\bf r},\sigma)\,
\exp\left(i\chi({\bf r})\right), \nonumber\\
\psi^{\dagger}({\bf r},\sigma) &\to& \psi^{\dagger}({\bf r},\sigma)\,
              \exp\left(-i\chi({\bf r})\right).
\label{calibr}
\end{eqnarray}
In this case the $1/q^2$ theorem of Bogoliubov for the Fermi
systems~\cite{Bogoliubov1} is valid which asserts that
in the presence of the pair condensate we have the inequality
for sufficiently small $q$
$$
\max\limits_{\omega,S} N_{\omega,{\bf q},S} \geq \frac{C}{q^2},
$$
where $N_{\omega,{\bf q},S}$ appears in Eq. (\ref{F2psiomq}) and
$\omega$ is the set of the quantum numbers corresponding to both
the continuous spectrum $(\omega=({\bf p},m_S))$ and the discrete
one $(\omega=i)$~\cite{Note8}. However, Eq. (\ref{Npqsms}) results
in
$$
N_{{\bf p},{\bf q},S,m_S} \leq 1
$$
because $n(k) \leq 1$ for fermions. Therefore, we have the only
possibility at which the singularity $1/q^2$ appears due to the
noncondensed bound pairs. It is reasonable to expect that these
pairs have the quantum numbers of the condensate couples $S_0,i_0$:
\begin{eqnarray}
\widetilde w_{S_0,i_0}({\bf q}) \geq \frac{C'}{q^2}.
\label{wi0q}
\end{eqnarray}

The BCS--model is not locally gauge invariant which results in
absence of the noncondensed bound pairs: $\widetilde w_{S,i}
({\bf q})=0$. It is important to note in this connection that
the bound pair states beyond the condensate may play a noticeable
role in calculating the gauge--invariant response of the system
to the electromagnetic fields.

We have proved that the noncondensed bound pairs coexist with the
condensed ones at $T < T_c$. So, any theory ignoring the
noncondensed bound pairs of fermions is not fully consistent.
Remark that the distribution of the bound fermion pairs over the
center--of--mass momenta obeys the inequality (\ref{wi0q}) with
$C'\propto k_B T n_0$ (see Ref.~\cite{Bogoliubov1}). The distribution
of the particles over momenta in the Bose gas $w(q)=n(q)/(2\pi)^3$
answers, at small $q$, the similar relation $w(q) \geq C''/q^2$
with $C''\propto k_B T n_0$ (here $n_0$ denotes the density of
the condensed bosons)~\cite{Bogoliubov1}. Therefore, there are
fundamental parallels between the Bose gas and the considered
subsystem of the fermion bound pairs. And these parallels are not only
reduced to agreement between the fermion--pair statistics and the
Bose one. Following this analogy, we can expect that the bound
fermion pairs exist even at $T > T_c$ (apparently, in some
temperature interval $T_c < T < T^*$, in spite of the disappearance
of the $1/q^2$--singularity). Thus, it looks as if any
superconducting phase transition is a particular case of the
Bose--Einstein condensation. This conclusion can be of interest
in the context of the discussion concerning different approaches
of investigating the high--$T_c$ superconductivity (see Refs.
\cite{Chakraverty,Alexandrov1}). Remark that possible experimental
consequences of the existence of fermion bound pairs beyond the
condensate can be found in paper\cite{Leggett} in the case of
neutral Fermi systems.

The space--uniform character of the Fermi system is of
use in the proof given above. Electrons in the crystalline field,
of course, can not be treated on the same level. However, for
$q \to 0$ (large wave lengths) a crystalline lattice can
be considered as continuum. Therefore, the derived result remains
correct in this case.

Emphasize that the bound pair states can fully be a result of the
collective effects. Indeed, as it was demonstrated by
Cooper~\cite{Cooper}, an arbitrary small attraction between electrons
leads to forming the condensate of the bound electron pairs. Hence,
if we considered a sufficiently shallow well as the two--fermion
interaction potential, we would observe formation of the condensed
and, according to the obtained result, noncondensed pairs at low
temperatures. However, the well can be chosen in such a way as to
prevent the bound states of two 'bare' fermions from appearing within
the ordinary two--particle problem.

At last, it is important to make one more remark on the connection
between the $1/q^2$ theorem of Bogoliubov and the Goldstone
theorem~\cite{Goldstone}. As it has been demonstrated in Ref.
\cite{Bogoliubov1}, existence of the Goldstone mode in the Bose
system results from the Bogoliubov theorem provided the mass
operator $\Sigma(\omega,k)$ is regular in the vicinity of the point
$\omega=0,\, k=0$. Let us emphasize that there are situations when
the Bogoliubov theorem is valid while it is not the case for the
Goldstone one. For example, in the case of neutral weakly
interacting Bose gas the condition mentioned above for the mass
operator is correct, and the Goldstone mode exists. On the contrary,
for the charged Bose gas the mass operator is not regular at $k=0$,
and, thus, there is no Goldstone mode. The similar situation is
realized for the Fermi systems (see, e.g. Ref.\cite{Wagner}).

\section{Conclusion}

\label{5}

Concluding, let us take notice of the main results once more. The
reduced density matrix of the second order is a fundamental
characteristic of a many--particle system, its eigenfunctions being
the pure states of two particles selected in an arbitrary way.
Appearance of the condensate of the bound pair states (\ref{n0})
implies the occurrence of the $\delta-$functional term in the
distribution of the bound pairs over the momentum of the couple
center of mass ${\bf q}$ (see Eq. (\ref{wiq})). Using the space
homogeneity of the system and the local gauge invariance
(\ref{calibr}) of the fermion interaction, we have proved that there
is the $1/q^2$-singularity in the distribution function $\widetilde
w_{S,i}({\bf q})$ provided that $n_0\not=0$. Thus, we refined the
$1/q^2$ theorem of Bogoliubov, having proved the singularity to
appear in $\widetilde w_{S,i}({\bf q})$. Therefore, presence of the
noncondensed bound pairs below $T_c$ is the necessary condition of
superconductivity.

A new simple proof of the Yang inequality for the Fermi systems
(\ref{N0N}) and its stronger variant (\ref{N0Dnk}) have also been
derived as results of secondary importance.

This work was supported by the RFBR Grant No. 97-02-16705.
Discussions with V. V. Kabanov and V. B. Priezzhev are gratefully
acknowledged.

\end{multicols}


\begin{references}
\bibitem{Randeria} M. Randeria, E-print {\tt cond-mat/9710223} (unpublished).
\bibitem{Anderson} P. W. Anderson, Phys. World {\bf 8}, 37 (1995).
\bibitem{Mott} N. F. Mott, Phys. World {\bf 9}, 16 (1996).
\bibitem{Baskaron} G. Baskaron, Z. Zou and P. W. Anderson, Solid St.
Comm. {\bf 63}, 973 (1987).
\bibitem{Alexandrov} A. S. Alexandrov and N. F. Mott, Rep. Progr. Phys.
{\bf 57}, 1197 (1994); A. S. Alexandrov, V. V. Kabanov and N. F. Mott,
Phys. Rev. Lett. {\bf 77}, 4796 (1996).
\bibitem{Blatt} J. M. Blatt, {\it Theory of Superconductivity} (New York
-- London, Acad. Press, 1964).
\bibitem{Bogoliubov1} N. N.~Bogoliubov, {\it Quasi--averages}, preprint
D--781, JINR, Dubna (1961) [English transl. N. N.~Bogoliubov, {\it
Lectures on Quantum Statistics}, vol. 2 (New York, Gordon and
Breach, 1970), p. 1].
\bibitem{Landau} L. D. Landau and E. M. Lifshitz, {\it Course of
Theoretical Physics}, vol. 3, {\it Quantum Mechanics -- Non--relativistic
Theory} (New York, Pergamon Press, 1977), \S 14.
\bibitem{Note1} It is more correct to speak about the bound pair
states rather than about bound pairs. Indeed, a particle
can form a bound state together with $M\,(M \geq 1)$ particles,
while it forms 'dissociated' states with the other $N-M$
particles.
\bibitem{Note9} Strictly speaking, this is valid for the $s$--wave pairing.
In general case the gap in the single--particle spectrum becomes ${\bf
k}$--dependent, one usually suppose that $\Delta=\Delta({\bf k})$ is
proportional to 'anomalous' averages $\langle a_{-{\bf k},\alpha}a_{{\bf
k},\beta}\rangle$. We associate the 'anomalous' averages with the wave
function of pairs in the condensate (see Eq.(\ref{phicond})).  As to the
exact relation between the binding energy $\varepsilon_{b}$ and the gap
$\Delta({\bf k})$ in general case, it is rather complicated question.
However, at any rate, the presence of the gap implies that
$\varepsilon_{b}\not=0$.
\bibitem{Note2} The continuous distribution can only be introduced
in the thermodynamic limit $N/V=\mbox{const}, \;V \to \infty$.
\bibitem{Leggett} A. J. Leggett, in {\it Modern Trends in the Theory
of Condensed Matter}, ed. by A. Pekalski and J. Przystawa
(Springer--Verlag, Berlin, 1980).
\bibitem{Schrieffer} J. R. Schrieffer, {\it Theory of
Superconductivity} (New York, Benjamin, 1964).
\bibitem{Note3} To consider the Gibbs grand canonical ensemble
it is sufficient to replace $\widehat H$ by $\widehat H -\mu
\widehat N$.
\bibitem{Bogoliubov2} N. N.~Bogoliubov, {\it Lectures on Quantum
Statistics}, vol. 1 (New York, Gordon and Breach, 1967) p. 39.
\bibitem{Note3a} After the thermodynamic limit $V \to \infty$.
\bibitem{Cherny} A. Yu. Cherny, E-print {\tt cond-mat/9807120},
submitted to Phys. Rev. A.
\bibitem{Note4} Of course, there is no magnetic ordering in the
superconducting state.
\bibitem{Bogoliubov3} N. N.~Bogoliubov, preprint
P--511, JINR, Dubna (1960) [English transl. in
Ref.~\cite{Bogoliubov1}, p. 76].
\bibitem{Bloch} C.~Bloch and C.~De Dominicis, Nucl. Phys. {\bf 7},
 459 (1958).
\bibitem{Note5} Eq. $\langle \psi(x)\rangle\not=0$ implies that there is the
condensate of the one--particle states which is impossible due to
the Pauli principle.
\bibitem{Note6} The Riemann's theorem asserts that for any regular
function $f({\bf q})$ (i.e. $f({\bf q})$ does not contain 
$\delta$--function) we have
$$
\lim_{r \to \infty}\int d^3 q\, f({\bf q}) \exp(i{\bf q}{\bf r})=0,
$$
provided the integral $\int d^3 q\, f({\bf q}) \exp(i{\bf q}{\bf r})$
exists.
\bibitem{Yang} C. N. Yang, Rev. Mod. Phys. {\bf 34}, 694 (1962).
\bibitem{Note7} Emphasize that some other
fields can also be included into the Hamiltonian, for example, the
phonon one. Thus, the Fr\"{o}hlich model is invariant with respect to
the transformation (\ref{calibr}).
\bibitem{Note8} Bogoliubov proved the theorem in the particular
case of the $s$--wave pairing, i.e. when in (\ref{phiconda})
$\varphi_s({\bf r})=\varphi(r)$, where $\varphi(r)$ is radially
symmetric function. The proof can easily be extended to the more
general case.
\bibitem{Chakraverty} B. K. Chakraverty, J. Ranninger, D. Feinberg,
Phys. Rev. Lett. {\bf 81}, 433 (1998).
\bibitem{Alexandrov1} A. S. Alexandrov, E-print {\tt
cond-mat/9807185}.
\bibitem{Cooper} L. N. Cooper, Phys. Rev. {\bf 104}, 1189 (1956).
\bibitem{Goldstone} J. Goldstone, Nuovo Cim. {\bf 19}, 154 (1961).
\bibitem{Wagner} H. Wagner, Z. Phys. {\bf 195}, 273 (1966).
\end{references}
\end{document}